\documentclass[final,journal,letterpaper]{IEEEtran}

 \usepackage[hidelinks]{hyperref}
 \usepackage{varioref} 
 \usepackage{wrapfig} 
 \usepackage{threeparttable} 
 \usepackage{dcolumn} 
  \newcolumntype{d}{D{.}{.}{-1}}
 \usepackage{nomencl} 
  \makeglossary
 \usepackage{subfigmat} 
 \usepackage{fancyvrb} 
  \fvset{fontsize=\footnotesize,xleftmargin=2em}
 \usepackage{lettrine} 
 \usepackage{algorithm,algorithmic}

 \usepackage{epsfig}
 \usepackage{epstopdf}
 \usepackage{subfigure}
 \usepackage{latexsym}
 \usepackage{amsmath}
 \usepackage{amsfonts}
 \usepackage{amssymb}
 \usepackage{mathrsfs}
 \usepackage{cases}
 \usepackage{array}
 \usepackage{color}
 \usepackage[table]{xcolor}
 \usepackage{soul}
 \setulcolor{red}
 \sethlcolor{yellow}

 \usepackage{url}
 \usepackage[noadjust]{cite}
 \usepackage{graphicx}
 \usepackage{psfrag}
 \usepackage{color}
   \usepackage{multirow}
 \usepackage{ragged2e}
 \usepackage{graphicx}
 \usepackage{RL_notation}
\usepackage{makecell}
\usepackage{titlesec} 
\usepackage{indentfirst} 
\usepackage{tabularx}

\newfont{\Bb}{msbm10 scaled\magstep1}

\titlespacing*{\subsection}
{0pt}{1.5ex minus .2ex}{1ex minus .2ex}
\titlespacing*{\subsubsection}
{0pt}{1 ex minus .2ex}{0pt}
\begin{document}
%
\title{Physics-informed Evolutionary Strategy based Control for Mitigating Delayed Voltage Recovery}
%
%
%

\author{Yan Du, Qiuhua Huang, Renke Huang, Tianzhixi Yin,
        Jie Tan, Wenhao Yu, Xinya Li
       

\thanks{The Pacific Northwest National Laboratory is operated by Battelle for the U.S. Department of Energy (DOE) under Contract DE-AC05-76RL01830. This work was supported by DOE ARPA-E OPEN 2018 Program. \newline
$\ast$ \textit{Corresponding authors: Qiuhua Huang, Renke Huang}}
\thanks{ Y. Du, Q. Huang, R. Huang, T.Yin, X.Li are with Pacific Northwest National Laboratory, Richland, WA 99354, USA (e-mail: \{yan.du, qiuhua.huang, renke.huang, tianzhixi.yin, xinya.li\}@pnnl.gov).}
\thanks{J. Tan, W. Yu are with Google Brain, Google Inc, Mountain View, CA, 94043 USA (e-mail: \{jietan, magicmelon\}@google.com).}}

\maketitle

\begin{abstract}
In this work we propose a novel data-driven, real-time power system voltage  control method based on the physics-informed guided meta evolutionary strategy (ES).
The main objective is to quickly provide an adaptive control strategy to mitigate the fault-induced delayed voltage recovery (FIDVR) problem.
Reinforcement learning methods have been developed for the same or similar challenging control problems, but they suffer from training inefficiency and lack of robustness for ``corner or unseen'' scenarios. On the other hand, extensive physical knowledge has been developed in power systems but little has been leveraged in learning-based approaches.  
To address these challenges, we introduce the trainable action mask technique for flexibly embedding physical knowledge into RL models to rule out unnecessary or unfavorable actions, and achieve notable improvements in sample efficiency, control performance and robustness.
Furthermore, our method leverages past learning experience to derive surrogate gradient to guide and accelerate the exploration process in training.
Case studies on the IEEE 300-bus system and comparisons with other state-of-the-art benchmark methods demonstrate effectiveness and advantages of our method.
\end{abstract}


%
\IEEEpeerreviewmaketitle

\section{Introduction}

%
%
%
%

\subsection{Motivation}
\IEEEPARstart{I}{NITIALLY} brought into the spotlight by the unprecedented success of AlphaGo in year 2016, the deep reinforcement learning (deep RL) technique \cite{sutton2018_RLBook} has been motivating breakthroughs in a broad range of areas including games, robotics, and autonomous driving. In the field of power systems, the deep RL technique has been leveraged for solving complex grid control and optimization problems, such as autonomous voltage regulation \cite{wang2020adata}, residential HVAC control \cite{yu2021multi}, electricity market bidding \cite{liang2020agent}, and power system stability and emergency control \cite{yan2018data,Huang2020_DRL,zhang2020deep}. The major advantages of the deep RL method over the conventional model-based method, as has been discussed thoroughly in the above existing research works, lie in that it is model-free and thus more robust than conventional control methods for modeling errors; it can determine control solutions within a very short time and meet real time control requirements; and it has generalization to unseen instances.

Nevertheless, there still exist a number of critical factors that prohibit the full adoption of  deep RL algorithms in the physical systems such as power systems: 1) they usually have a costly training and fine-tuning process due to numerous embedded parameters and low exploration efficiency; 2) they fail to properly incorporate physical knowledge to achieve efficient training and robust performance; 3) they cannot adapt well to new or unseen situations. 

The objective of this paper is to address these three key issues by embedding the physics knowledge for accelerating the learning process and improving the robustness of the control policy, by incorporating guided search to achieve better exploration efficiency, and by leveraging meta-learning for fast adaptation.

Physics-informed machine learning has received growing attention lately due to the challenges encountered by the pure data-driven machine learning methods, including high cost of data acquisition, data incompleteness, and extremely high search spaces, etc \cite{raissi2017physics}.
In the fault-induced delayed voltage recovery (FIDVR) problem, the bulk power system with a large number of load buses induces a vast control action space and an unduly burdensome searching process. Voltage performance criteria have been developed by the industry through extensive off-line studies to guide planning and operation against voltage problems including FIDVR. 
In light of these, we augment the conventional RL model (represented by a neural network) with a physics-informed module called trainable action mask (TAM), which utilizes power system physical knowledge (i.e., voltage performance criteria) to filter out improper control actions and to avoid unnecessary explorations, thereby achieving better sample efficiency and control robustness. A recent study in the dialog system \cite{wu2019tam} adopted a similar idea to leverage prior, non-physical knowledge. In this paper, we for the first time embed physical knowledge into RL models through the TAM technique for power system control applications, particularly, the FIDVR problem and got promising results. 

Conventional deep RL methods rely on chain-rules and back-propagation to update the parameters of the neural networks, which makes it rather difficult to incorporate non-differentiatable modules like TAM. Recently, evolutionary strategy (ES) has been proven to be an effective, scalable alternative to conventional RL methods \cite{salimans2017evolution,mania2018simple}. ES methods are derivative-free and can be easily parallelized. The derivative-free feature facilitates incorporating the TAM technique into the RL model.

Most recently, a novel guided ES method is proposed to enhance the exploration efficiency of the algorithm in high-dimensional parameter spaces \cite{maheswaranathan2019guided}. 
Instead of conducting a complete random search, the guided ES method leverages the guidance from a surrogate gradient, which is derived from prior exploration experiences during training. By coordinating the random search with the gradient-guided search, the guided ES method can achieve a faster learning speed and better solutions.

Inspired by the above work, also building upon our previous work in leveraging an advanced ES algorithm for power system emergency control \cite{huang2021accelerated_DRL}, in this paper we develop a model-free guided ES-based control strategy to mitigate FIDVR with high exploration efficiency. 
The free from computationally intensive back-propagation process and easy parallelization of the guided ES method makes it possible to overcome the extremely high-dimensional state/action complexities introduced by large-scale power systems.

One key issue with the FIDVR problem is that how well a developed control policy can be adapted to the ever-changing grid operation conditions \cite{park2020model}.
To achieve a fast adaptation of the learnt control strategy to unseen fault scenarios to meet with the real-time control requirement, we further combine the above guided ES method with a meta-learning strategy introduced in our previous work \cite{huang2021learning}, namely the meta strategy optimization (MSO), which leads to the guided meta ES algorithm. The core idea behind MSO is to learn a latent variable as a representation of the variations of the training environments. The latent variable can then be fine-tuned when a new operation scenario is presented, and the control policy is adjusted accordingly. 
The adaptability of the proposed guided meta ES method makes it practical for real-world applications.

\subsection{Literature review}

Our work focus on the FIDVR problem. FIDVR is defined as the phenomenon whereby system voltage remains at significantly reduced levels for several seconds after a fault has been cleared \cite{NERC2009}. The root cause is stalling of residential air-conditioner (A/C) motors absorbing excessive reactive power from the power grid and prolonged tripping. FIDVR events occurred in many utilities in the US. Concerns over FIDVR issues have increased since residential A/C penetration is at an all-time high and growing rapidly. A transient voltage recovery criterion (TVRC) is defined to evaluate the system voltage recovery. Without loss of generality, we referred to the standard proposed in \cite{PJM2009} and shown in Fig. \ref{fig:voltage_th}. After fault clearance, the standard requires that voltages should return to at least 0.8, 0.9 and 0.95 p.u. within 0.33 s, 0.5 s and 1.5 s, respectively. 

The control actions to mitigate the FIDVR problem can be classified as supply-side solutions and demand-side solutions. The most popular supply-side solution applied by utilities is the implementation
of FACTS devices such as SVC \cite{al2009preventing} and static condenser\cite{du2009utilizing} to increase reactive power support. However, these devices are very expensive and cost approximately \$20–50 million per installation. Since a large number of these devices would have to be installed, it could be very costly. The demand-side load-shedding solution is widely recognized as the most effective and economic control strategy to mitigate the FIDVR problem by shedding part of the stalling A/C motors in the impacted areas to reduce the huge amount of reactive power they are absorbing and thus recover the voltage \cite{Bai2011}. The challenges associated with determine the load shedding actions to mitigate the FIDVR problem are when, where, and how much amount of A/C motor load to shed in the impacted areas of the grid to recover the system voltage. It is desired that shedding the least amount of the A/C motor load in the system to recover the system voltage to meet the transient voltage recovery criterion.

Existing load shedding methods for mitigating the FIDVR problem can be roughly classified into the following four categories, and the advantages of our proposed method over each category are analyzed accordingly:  
\begin{itemize}
    \item \textit{rule-based methods}: An example of rule-based method can be found in \cite{lefebvre2003design}. While the implementation is fairly easy, the settings of the rules tend to be conservative, and they cannot adapt flexibly to different conditions. 
    \item \textit{model-based methods}: one popular method is the model-predictive control (MPC) method \cite{Jin2010MPC,MPC2}. The MPC model can be formulated based on the detailed power grid network and dynamic models, and can be solved based on the mathematical-rigor optimization techniques. However, inclusion of the transient stability constraints in the FIDVR problem makes the MPC model much more complex and higher dimensional when compared with the conventional steady-state OPF problem. Thus, it suffers from the scalability issue and cannot not meet the solution time requirement (i.e. \textless0.5 s ) for real-time voltage control.
    \item \textit{measure-based methods}: new real-time voltage control methods have been developed by leveraging the phasor measurement unit (PMU) technologies \cite{glavic2012see,matavalam2019pmu,sun2019review}. They do not consider coordinating load shedding actions to achieve system-wide optimality.
    \item \textit{learning-based methods}: the deep Q network (DQN) and deep deterministic policy gradient (DDPG) methods have been applied for developing emergency load shedding schemes to recover voltage \cite{Huang2020_DRL,zhang2018load}. While the deep RL methods have been credited for their strong exploration and adaptability, they do suffer from training inefficiency and lack of adaptability and robustness, which prohibits their deployment for large-scale power system voltage recovery.
\end{itemize}
Compared with the above existing methods, our proposed evolutionary strategy method has the following merits in the case of real-time FIDVR mitigation control:
\begin{itemize}
\item Compared with the rule-based method, our proposed method adopts the neural network to learn a more generalized control strategy, and further combines with a meta learning strategy to get quickly adapted to unseen operation scenarios and contingencies.
\item Compared with the model-based method, our proposed method is model-free and once well-trained, can be directly executed in real-time, which greatly spares both modeling efforts and computational efforts.
\item Compared with the measurement-based method, our proposed method can efficiently tackle with high-dimensional state and action spaces in the case of large-scale power systems with the embedded neural network to achieve a near-optimal solution.
\item Compared with other learning-based deep RL methods, our proposed method has a much higher exploration efficiency and can be easily scaled up for parallel computing during the training, and furthermore our proposed method is guided by physical knowledge during both training and execution, which greatly reduces the training efforts and improves the control adaptability and robustness.
\end{itemize}
Table~\ref{table:contribution} presents an overview of the proposed technical roadmap in our work. We also analyze the improvement of the current work over our previous works in this research area \cite{huang2021accelerated_DRL,huang2021learning} as follows:
\begin{itemize}
    \item Compared with \cite{huang2021accelerated_DRL}, which applies a parallel evolutionary strategy method called augmented random search (ARS), we introduced guided search to the evolutionary strategy in this work instead of a complete random search, which substantially accelerates the learning process and help find better control strategies.
    \item Compared with \cite{huang2021learning}, which adopts the meta strategy optimization (MSO) to develop a robust and adaptive load shedding strategy, in this work we combine the MSO with the guided search to take advantage of both methods to improve the computational efficiency as well as the adaptability of the evolutionary strategy for real-time voltage recovery.
    \item Last but not least, we for the first time introduce physical knowledge into the model-free ES method through the TAM technique to help boost the exploration efficiency. This endeavor shows remarkable success yet was not explored in the previous two pieces of works.
\end{itemize}

\begin{table}[t]
\caption{Technical Roadmap of the Physics-informed ES Method}
\begin{center}
\begin{tabular}{|c|c|}
\hline
\textbf{Key challenges of deep RL}&\textbf{Proposed technique}\\
\hline
\multicolumn{1}{|p{4cm}|}{Costly back-propagation process; Laborious parameter fine-tuning; Lack of scalability; Increasing algorithm complexity}&\multicolumn{1}{p{4cm}|}{Derivative-free ES for easy paralllelization and low computational burden (in our previous work\cite{huang2021accelerated_DRL})}\\
\hline
\multicolumn{1}{|p{4cm}|}{Lack of adaptability and generalization to unseen test cases}&\multicolumn{1}{p{4cm}|}{Meta strategy optimization for fast policy adaptation (in our previous work \cite{huang2021learning})}\\
\hline
\multicolumn{1}{|p{4cm}}{Time-consuming random action-space exploration}&\multicolumn{1}{|p{4cm}|}{Guiding exploration in the parameter space with surrogate gradient to focus on promising directions (\textbf{developed in this paper})}\\
\hline
\multicolumn{1}{|p{4cm}}{High-dimensional action domains leading to exhaustive searching; lack of robustness}&\multicolumn{1}{|p{4cm}|}{Introducing physical knowledge through TAM for effective action filtering (\textbf{developed in this paper})}\\
\hline
\end{tabular}
\label{table:contribution}
\end{center}
\vspace{0mm}
\end{table}

\subsection{Contributions}
In summary, the key contributions of our work can be outlined as follows:

1) The major contribution of our work is the novel embedding of physical knowledge into the model-free ES method to achieve high exploration efficiency for the real-time FIDVR problem with large action domains. The physical knowledge is introduced through the TAM technique, which eliminates improper load shedding actions and considerably spares the exploration efforts. To the best of the authors’ knowledge, this effort of embedding physics awareness in model-free voltage control methods is unprecedented in the literature. We have witnessed substantial algorithm performance improvement in our comparative studies.

2) A novel model-free guided meta ES method is developed to achieve superior sample efficiency than the standard ES algorithm, thanks to the combination of random search with surrogate gradient information. 

3) The generalization and adaptability of the learnt control policy to unseen fault scenarios is further enhanced through a meta learning strategy.  


4) The adaptability and efficiency of the proposed voltage recovery method based on the physics-informed guided meta ES algorithm is fully verified  by testing on a large-scale power system under multiple unseen fault scenarios and by comparing with state-of-the-art benchmark methods, which implies its great promises for real-world applications.


\subsection{Organization}
The rest of the paper is organized as follows: Section II describes the problem formulation of FIDVR. Section III introduces the proposed adaptive model-free control method based on guided meta ES algorithm. In Section IV, the physics knowledge is further combined with the proposed control method through trainable action mask.
Case studies are shown in Section V. 
Finally, Section VI concludes the work and the future directions.

\section{Problem Formulation}
\begin{figure}[t]
\centerline{\includegraphics[scale=0.45]{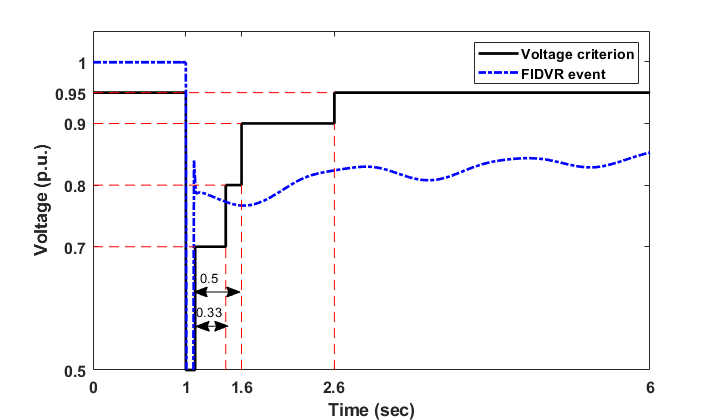}}
\caption{Transient voltage recovery criterion}
\label{fig:voltage_th}
\end{figure}

As discussed in Section I.B, one widely-recognized, effective and economic control strategy to mitigate the FIDVR problem is load shedding. An ideal load shedding strategy should be able to bring the system voltage magnitude to a certain level with minimum amount of load shedding.
A standard transient voltage recovery criterion is shown in Fig. \ref{fig:voltage_th} \cite{PJM2009}. As shown in the figure, the voltage should return to at least 0.7, 0.8, and 0.9 p.u. within 0.33s, 0.5s and 1.5s after the fault is cleared.
Deciding the optimal load shedding strategy is not a trivial task since three crucial problems must be considered: when to conduct load shedding, at which bus the load should be shed, and how much of the load should be shed, which leads to a high-dimensional non-convex decision-making problem \cite{Huang2020_DRL} and fails many model-based solutions in the case of real-time implementation. Also, the model-based solution is sensitive to the model errors which are common in complex power systems.

Based on the above discussions, in this work we propose to formulate FIDVR mitigation control problem as a Markov Decision Process (MDP) and further apply a model-free guided meta ES algorithm with trainable action mask (TAM) to obtain the optimal load shedding strategies under different fault scenarios, which is adaptive, highly scalable, and computationally efficient. The MDP-based problem formulation is first presented as follows.

The MDP is represented as a 5-tuple ($\mathcal{S,A,P},r,\gamma$), and their definitions under the context of FIDVR are provided as follows:

\noindent1) \textbf{state}: 
the state $\mathcal{S}$ is defined as a vector that contains the latest observations from the power system, including the voltage magnitude and the percentage of remaining load that can be shed at the monitored buses: $s_{t} = [V_{1,t},...,V_{M,t},P^{d}_{1,t},...,P^{d}_{N,t}]$, where $M$ and $N$ are the number of voltage monitoring buses and controllable load buses.

\noindent2) \textbf{action}: the action $\mathcal{A}$ is defined as a vector that contains the normalized load shedding actions for all the controllable buses. The normalized load shedding action is a scalar between -1 and 1, where -1 indicates that 20\% of the remaining load will be shed, and 1 indicates no load shedding actions.

\noindent3) \textbf{state transition}: the state transition $\mathcal{P}$ describes the power system dynamics and is deterministically governed by a set of differential and algebraic equations:
\begin{equation}
    \dot{x}_t= f(x_{t},y_{t},d_{t},a_{t})
    \label{eq:differential}
\end{equation}
\vspace{-5mm} 
\begin{equation}
     0=g(x_{t},y_{t},d_{t},a_{t})
    \label{eq:algebraic}
\end{equation}
In (\ref{eq:differential})-(\ref{eq:algebraic}), $x_{t}$ represents the system dynamic state variables, such as the generator rotor angle and speed; $y_{t}$ represents the system algebraic state variables, which are usually bus voltage magnitudes and bus voltage angles; $d_{t}$ refers to the system perturbation or contingency; and $a_{t}$ is the control action.  

\noindent4) \textbf{reward}: in the power system emergency control problem, the main objective is to recover the voltage magnitude to the normal level after fault clearance with the least amount of load shedding. To reach this objective, we adopted the same reward design as in our previous work \cite{Huang2020_DRL}:
\begin{equation}
r_{t} = 
\begin{cases}
      -10000,\, if\,V_{i,t} < 0.95,\, T_{pf} + 4 < t\\
      c_{1}\sum_{i}\Delta V_{i,t} - c_{2}\sum_{j}\Delta P_{j,t}(p.u.) - c_{3}u_{ivld}, \,     \\
      \text{otherwise}
    \end{cases}
\label{eq:reward1}
\end{equation}
where
\begin{equation}
\Delta V_{i,t} = 
\begin{cases}
      \text{min}\,\left\{V_{i,t}\!-\! V_{th,1},0 \right\},\, if\,T_{pf} \!< \!t\!<\! T_{pf} \!+\! t_{1}\\
      \text{min}\,\left\{V_{i,t} \!-\! V_{th,2},0\right\},\, if\,T_{pf} \!+\! t_{1} \!<\! t \!<\! T_{pf} \!+\! t_{2}\\
      \text{min}\,\left\{V_{i,t} \!-\! V_{th,3},0\right\},\, if\,T_{pf} \!+\! t_{2} \!<\! t 
    \end{cases}
\label{eq:reward2}    
\end{equation}
In (\ref{eq:reward1})-(\ref{eq:reward2}), \textit{t} is the current time step;$V_{i,t}$ is the voltage magnitude at bus $i$;$T_{pf}$ is the time instant for fault clearance; $\Delta P_{j,t}$ is the amount of shed load at bus $j$ in p.u.; $u_{ivld}$ is the penalty for invalid action if there is still a load shedding action at buses with zero remaining load; $c_1$,$c_2$, and $c_3$ are the weight factors; $V_{th,1}$, $V_{th,2}$,$V_{th,3}$, $t_{1}$,and $t_{2}$ constitute the voltage recovery criterion. One example of their values has been shown in Fig.~\ref{fig:voltage_th}.  
\noindent5) \textbf{discount factor}: the objective of MDP is to maximize the following total reward $R_{t} = \sum_{t'=t}^{T}\gamma^{t'-t}r_{t'}$,where $\gamma$ is a discount factor between 0 and 1. The reason for adding the discounted factor is to avoid an infinite sum of future rewards.



    



\section{Guided Meta Evolutionary Strategy}
In this section, we will first give a brief review of the evolutionary strategy (ES). 
Then we will introduce the guided ES algorithm that combines the surrogate gradient with random search to achieve higher sampling efficiency. 
Lastly, we will present a guided meta ES algorithm by utilizing meta strategy optimization (MSO) to obtain adaptive control strategies.

\subsection{An introduction to ES}

The ES is a type of heuristic search algorithm inspired by the evolution theory: at each iteration, a population of parameters that need to be optimized are randomly perturbed and their objective function values are calculated. The parameters with the highest values are then recombined to formulate the population for the next iteration. The process repeats until the objective meets the convergence criterion.

In the context of RL, given the reward function $r$ and the policy $\pi(s|\theta)$, the goal is to find the optimal $\theta$ that maximizes the expected total discounted reward $E\{\sum^{T}_{t=0}\gamma^{t}r_{t}(s_{t},\pi(s_{t}|\theta))\}$. \textbf{Algorithm 1} shows the implementation of ES \cite{salimans2017evolution}:
\begin{algorithm}
\label{algo:ES}
 \caption{Evolutionary Strategy (ES)}
 \begin{algorithmic}[1]
  \STATE Initialize the learning rate $\eta$, noise standard deviation $\sigma$, the number of perturbation directions $N$, and policy parameter $\theta$
  \FOR {iteration $t$ = 1 \TO $M$}
     \STATE Sample perturbation directions $\epsilon_{1},...,\epsilon_{N}$ from $\cal{N}$(0,I)
     \FOR {$i$ = 1 \TO $N$}
         \STATE Generate action $a_i = \pi(s|\theta_{t} + \sigma\epsilon_{i})$ 
         \STATE Execute $a_i$ and receive reward $r_i$
     \ENDFOR
     \STATE Update the policy parameter:
     \STATE $\theta_{t+1} =\theta_{t} + \eta\frac{1}{N\sigma}\sum_{i=1}^{N}r_{i}\epsilon_{i}$
  \ENDFOR
\end{algorithmic} 
\end{algorithm}

As shown in the above pseudo code, the algorithm consists of two repeated phases: first, the policy parameter is randomly perturbed by noises derived from a standard normal distribution, and the associated actions are executed and evaluated based on their reward values for an entire episode (line 3-line 7); second, the policy parameter is updated by an estimated stochastic gradient (line 9), which comes from the following derivation: assuming our objective is to optimize $\theta$ over a distribution $p_{\psi}(\theta)$ to maximize the expected reward $\mathbb{E}_{\theta\sim p_{\psi}(\theta)}r(\theta)$, when the parameter distribution $p_{\psi}(\theta)$ follows a Guassian distribution, the expected reward can be directly written as $\mathbb{E}_{\theta\sim p_{\psi}(\theta)}r(\theta)\!=\!\mathbb{E}_{\epsilon \sim N(0,I)}r(\theta+\sigma\epsilon)$. With the objective defined in terms of $\theta$, the gradient can be calculated as follows:
\begin{equation}
    \nabla\mathbb{E}_{\epsilon \sim N(0,I)}r(\theta+\sigma\epsilon) = \frac{1}{\sigma}\mathbb{E}_{\epsilon \sim N(0,I)}\left\{r(\theta+\sigma\epsilon)\epsilon\right\}
    \label{eq:stochasticgradient}
\end{equation}

The expectation term in (\ref{eq:stochasticgradient}) can be achieved through sampling, as shown by line 3 in the algorithm.
Note that line 4-line 7 can be naturally deployed in a parallel fashion to speed up the training, since each perturbation direction $\epsilon_{i}$ is independent from each other.
The simple way of sampling instead of back-propagation for parameter update makes the ES algorithm more scalable than the gradient-based RL methods. We will provide more explanations on scalability with parallel computing in the context of power system control in the later subsection IV-B.

\subsection{Guiding ES search with surrogate gradient}
In the above ES algorithm, the policy parameter $\theta$ is randomly perturbed following a Gaussian distribution. While this random search is easy to implement, it can introduce high variance and results in unnecessary explorations. The guided ES algorithm is thus proposed to handle this challenge. The core idea behind the guided ES algorithm is to refer to the surrogate gradient to guide the algorithm search toward the most promising directions instead of conducting a completely random search.

\begin{figure}[t]
\centerline{\includegraphics[trim=0cm 1.5cm 2.2cm 3.0cm, clip,scale=0.3]{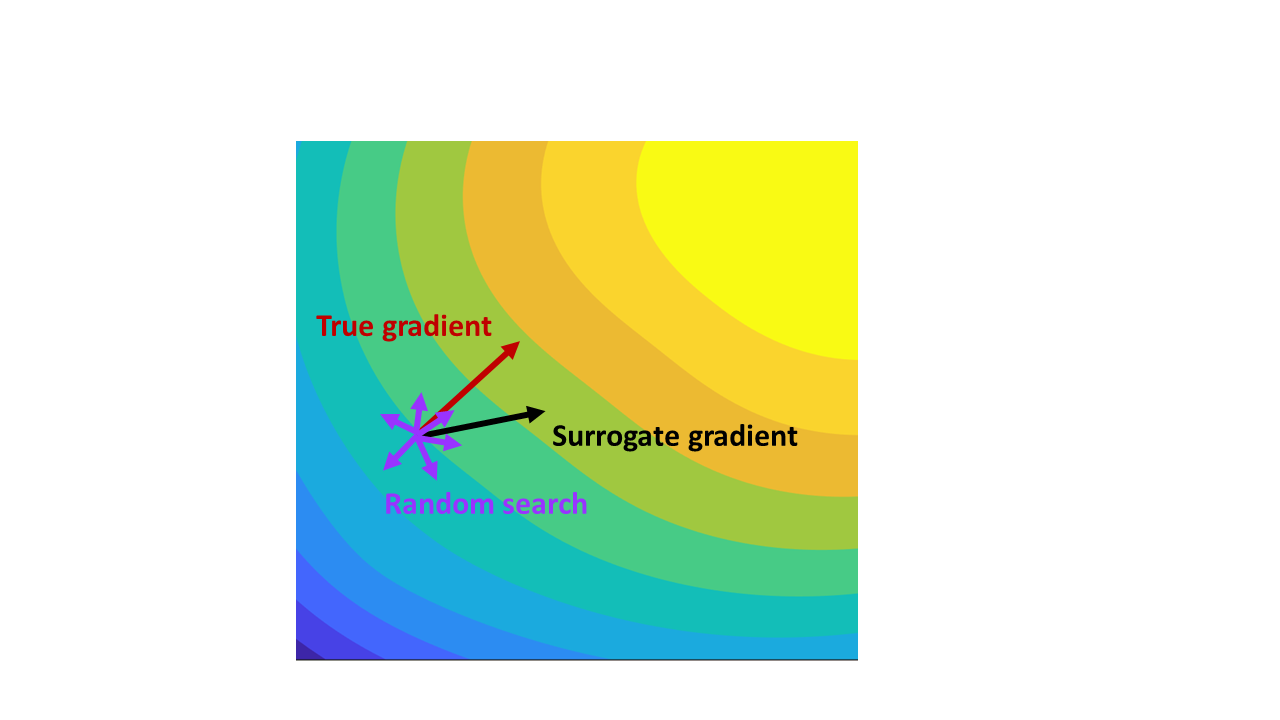}}
\caption{Schematic of surrogate gradient}
\label{fig:ESillustration}
\end{figure}

A surrogate gradient is correlated with the true gradient, but somehow biased or corrupted due to the model unobservability. An illustration of the surrogate gradient is shown in Fig.  \ref{fig:ESillustration}.
The guided ES algorithm takes advantage of the surrogate gradient in the following way \cite{maheswaranathan2019guided}: suppose we can get a vector of surrogate gradient for the policy parameters at each iteration, then by collecting the surrogate gradients from the previous $k$ iterations, we can generate a subspace $U^{T}U=I_{k}$, where $U$ is an $n\times k$ orthogonal basis for this subspace, and $n$ is the dimension of the policy parameters. The gradient information can be further embedded in the ES algorithm by changing the distribution of the perturbation $\epsilon_{i}$ from $\cal{N}$(0,I) to $\cal{N}$(0,$\sum$), where $\sum$ is calculated as follows: 
\begin{equation}
    \sum = \alpha^{2}I_{n} + (1-\alpha)^2UU^{T}
    \label{eq:noisedistribution}
\end{equation}

In (\ref{eq:noisedistribution}), $\alpha$ is a weight factor that makes a trade-off between the random search (exploration) and the guided search with surrogate gradient (exploitation). Setting $\alpha=1$ will lead to the ES algorithm. In our case, we set $\alpha$ to 0.5 to balance the exploration with exploitation. With the modified distribution, the perturbation direction $\epsilon_{i}$ can be calculated as follows:
\begin{equation}
    \epsilon_{i} = \alpha\epsilon^{\prime} + (1-\alpha)\epsilon''
    \label{eq:disturbance}
\end{equation}

where $\epsilon' \sim \cal{N}$(0,$I_n$), and $\epsilon'' \sim \cal{N}$(0,$I_k$). The complete guided ES algorithm is shown in \textbf{Algorithm 2}.
The algorithm basically follows the same framework as the ES method. 
One difference is that at the initialization state, a surrogate gradient buffer $B$ is defined to store the surrogate gradients from the previous $k$ steps for generating perturbations.
In addition, an antithetic sampling is applied, where for each perturbation direction, a pair of evaluations for $\theta_{t}+\sigma\epsilon_{i}$ and  $\theta_{t}-\sigma\epsilon_{i}$ are conducted to reduce variance, as shown by line 6. 
The evaluations are later used to calculate the surrogate gradient for policy parameter update, as shown by line 10-11.
Finally, the surrogate gradient is stored in the buffer $B$ for generating the new perturbation distribution, as shown by line 12 in the algorithm.
\begin{algorithm}
\label{algo:guided ES}
 \caption{Guided Evolutionary Strategy (guided ES)}
 \begin{algorithmic}[1]
  \STATE Initialize the learning rate $\eta$, the weight factor $\alpha$, the scale factor $\beta$, the noise standard deviation $\sigma$, the number of samples $N$, the number of surrogate gradients to use $k$, and the policy parameter $\theta$
  \STATE Initialize the surrogate gradient buffer $B\in\mathbb{R}^{k\times n}$
  \FOR {iteration $t$ = 1 \TO $M$}
    \STATE Sample\,perturbation\,directions $\epsilon_{i},...,\epsilon_{N}$ from $\cal{N}$(0,$\sum$)
     \FOR {$i$ = 1 \TO $N$}
         \STATE Generate action $a_i\!=\! \pi(s|\theta_{t}\!+\! \sigma\epsilon_{i})$,\,$a_i\!=\! \pi(s|\theta_{t}\!-\! \sigma\epsilon_{i})$
         \STATE Execute $a_i$ and receive reward $r_{i}$
    \ENDFOR
    \STATE Update $\theta$ with surrogate gradient $g$:
    \STATE $g\!=\!\frac{\beta}{2\sigma N}\sum_{i=1}^{N}\epsilon_{i}[r(\theta_{t}+\sigma\epsilon_{i})\!-\!r(\theta_{t}-\sigma\epsilon_{i})]$
    \STATE $\theta_{t+1}\!=\!\theta_{t} + \eta g$
    \STATE Store $g$ to the buffer $B$ and update the surrogate gradient subspace $U$ and perturbation distribution $\sum$
  \ENDFOR
\end{algorithmic} 
\end{algorithm}
\subsection{Enhancing algorithm adaptability with MSO}
The power system has a fast-changing and uncertain nature, which requires that an emergency control strategy should have sufficient robustness and be adaptive to unseen fault scenarios. To enhance the adaptability of the above data-driven guided ES-based control policy to new environment dynamics, in this subsection we propose to 
integrate the idea of meta learning, namely learning to learn, into the guided ES method, 
which leads to guided meta ES.

We apply a specific meta-learning technique, the meta strategy optimization (MSO) \cite{yu2020learning}, to realize the above objective.
MSO adapts a learnt control policy to unseen scenarios through latent space representation.
For each operation scenario encountered during the training, a latent variable is defined for this scenario to encode its hidden features. 
The latent variable is later combined with the direct observations of the scenario and sent to the policy function for decision-making.
The latent variable optimization and the policy parameter update can be expressed by the following two equations: 
\begin{equation}
    c_{\mu,t}=\mathop {\arg \max }\limits_c J_\mu(c,\theta_{t})
    \label{eq:maxc}
\end{equation}
\vspace{-.2cm}
\begin{equation}
    \theta_{t+1}=\mathop {\arg \max }\limits_\theta E_{\mu}[J_{\mu}(c_{\mu,t},\theta)]
    \label{eq:maxtheta}
\end{equation}

In (\ref{eq:maxc}), $c_{\mu,t}$ is the latent variable associated with scenario $\mu$ at the $t^{th}$ iteration; $J_\mu(c,\theta_{t})$ is a performance measurement, e.g., the reward function. 
The policy parameter $\theta$ is then optimized by maximizing the expected performance measurement with the learnt latent variable $c_{\mu,t}$, as shown by (\ref{eq:maxtheta}).

When unseen operation scenarios occur during the testing, new latent variables can be  calculated through the above process for fine-tuning the policy, making it adapted to the new environment dynamics.
This adaptation can be realized through only a few iterations with the environment, which is highly time-efficient.
More technical details of MSO application in power system emergency control can be found in our previous work \cite{huang2021learning}.



\subsection{Advantages of Guided Meta ES algorithm over conventional RL}
Under the context of power system FIDVR mitigation control, the above guided meta ES algorithm exceeds the gradient-based RL methods in the following three aspects:

\begin{itemize}
\item The evolutionary strategy does not require or rely on back-propagation process for gradient calculation and parameter update. This provides more flexibility such as incorporating a trainable action mask module, which can be regarded as a non-differentiatable regulation layer,in the end-to-end training process.
\item In line with the above analysis, the large-scale power system FIDVR mitigation control problems have the feature of non-smoothness in the environment dynamics and the reward function definition, which can cause the issue of gradient explosion during the back-propagation process in standard value-based and policy gradient deep reinforcement methods. The proposed evolutionary strategy refers to the surrogate gradient as an efficient workaround to avoid the gradient explosion issue.
\item The evolutionary strategy is well suited to scale up with parallel computing technologies: the algorithm operates on complete power system dynamic simulations, which indicates infrequent communications among the parallel workers. Considering the variety of power system operation scenarios, a parallel simulation greatly facilitates the training process.
\item In the case of unseen  scenarios, the learnt control policy can be quickly adjusted through MSO for new environment dynamics, which is highly desirable for real-time FIDVR mitigation under uncertainties.
\item The learning performance of the standard value-based and policy gradient deep RL methods are highly sensitive to the hyper-parameter settings, and they require additional human efforts for parameter fine-tuning. In contrast, the proposed evolutionary strategy method only has a small set of hyper-parameters, which avoids extensive fine-tuning process while boosting a better control performance.
\end{itemize}


\section{Physics-informed Guided Meta ES with Trainable Action Mask}
In this section, we aim to further improve the exploration efficiency of the guided meta ES method by introducing a novel trainable action mask (TAM) technique, which brings in the physical knowledge of power systems to pinpoint the optimal control actions. 
\subsection{Embedding physics knowledge through TAM}
While the guided ES algorithm is much better than basic ES algorithms, it still suffers from exploration inefficiency issues when applied to high-dimensional control problems.
One way to overcome this obstacle is to incorporate a physics-informed action mask component into the algorithm, which will filter out impossible or unfavorable actions and prevent the algorithm from conducting unnecessary explorations \cite{williams2017hybrid}.

\begin{figure}[t]
\centerline{\includegraphics[scale=0.40]{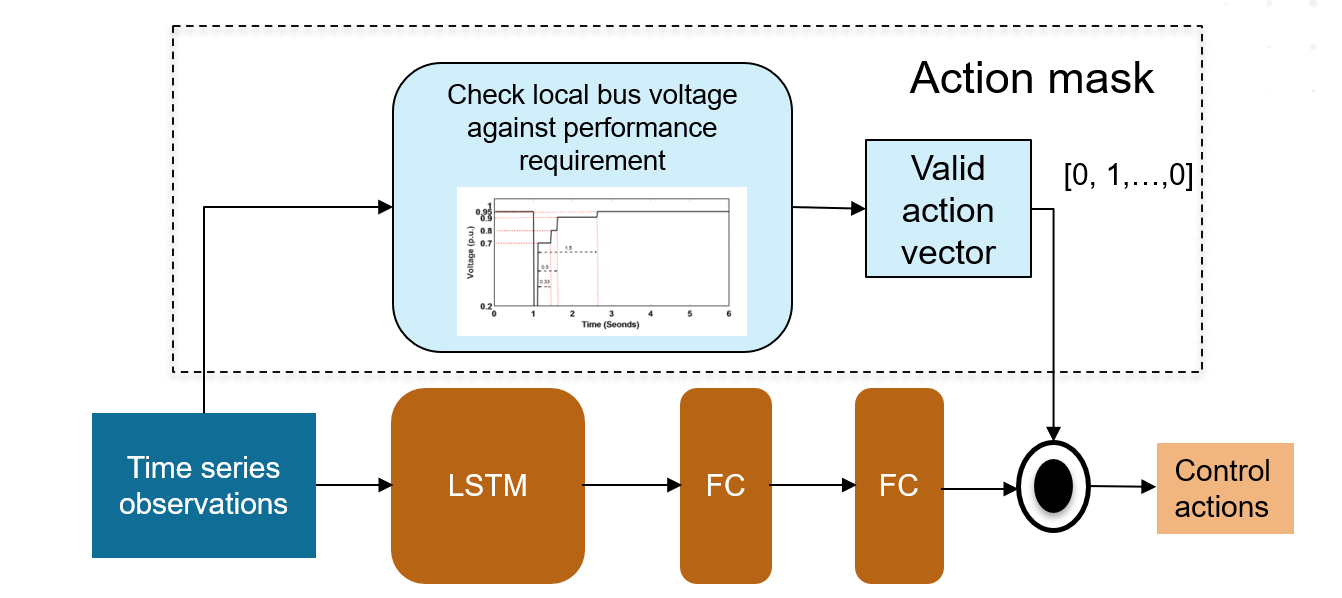}}
\caption{Regulating the neural network output with a fixed action mask}
\label{fig:fixed_action_mask}
\end{figure}

The action mask makes use of existing physical knowledge.
In the case of FIDVR mitigation, the voltage performance criterion (such as the one in Fig.~\ref{fig:voltage_th}) can be regarded as prior knowledge, and be used to accelerate the training and improve control performance through a simple hand-crafted action-mask, as illustrated in Fig. \ref{fig:fixed_action_mask}. In the figure, the time-sequential observations first go through a long-short-term-memory (LSTM) layer, then two fully-connected (FC) layers. The output from the FC layer is later masked by an action vector to get the final control actions. The function of the LSTM layer is to learn the temporal correlations of the system observations over a long time window. In the LSTM layer, a cell state is designed to capture and maintain the historical input, which spares the efforts of stacking all the historical data as one single input and thus reduces the dimension of the input data.

The action mask is constructed as follows:
The action mask has the same dimension as the control action.
At each time step, for each controllable bus, if its observed voltage magnitude is above the stability 
criterion, no action is required and a zero element will be added to the corresponding position in the mask, and vice versa.
Next, the action generated by the policy network will be multiplied by this mask, where the positions with zero elements will eliminate the corresponding load shedding actions since it is unnecessary, and the positions with one will keep the load shedding actions.

Note that in the above hand-crafted action mask, the mask settings are set according to a predefined, fixed performance or stability criterion and generally remain fixed for all scenarios.
However, considering that the power system operation scenarios can vary significantly from one to another, for instance, with different loading conditions, a fixed mask is unlikely to be the optimal solution for a wide range of operation scenarios. 

\begin{figure}[t]
\centerline{\includegraphics[scale=0.40]{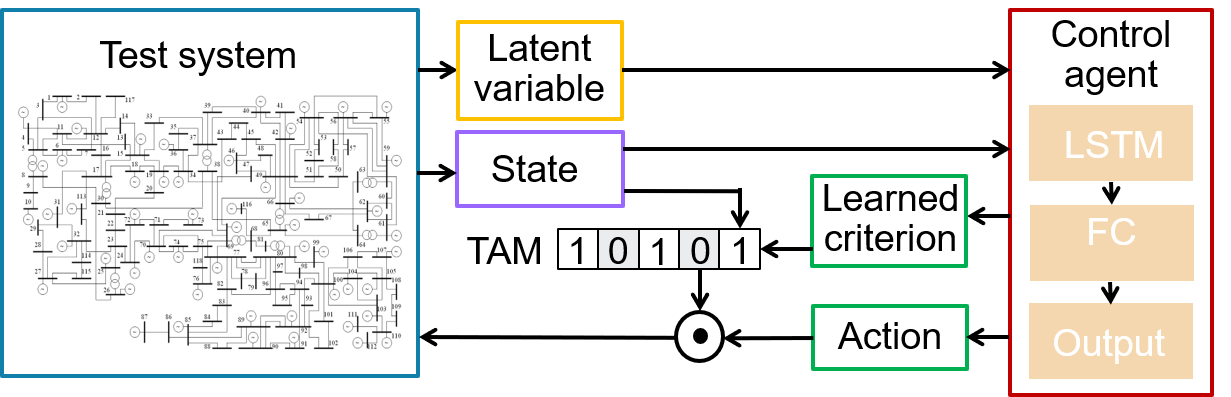}}
\caption{Illustration of the trainable action mask technique}
\label{fig:actionmask}
\end{figure}

Based on the above discussions, we propose a TAM technique to develop a learnable, adaptive criterion and to obtain a more flexible and generalized control strategy.
An illustrative explanation of the TAM technique is shown in Fig.~\ref{fig:actionmask}, and it can be described by the following mathematical expressions:
\begin{equation}
    [a_{t},cr_{t}]= \pi(s_{t},c_{t}|\theta_{t})
    \label{eq:policyoutput}
\end{equation}
\vspace{-.5cm}
\begin{equation}
    TAM_{i,t} = 
\begin{cases}
      \text 1,\, if\,s_{i,t} \!< cr_{t}\\
      \text 0,\, \, elsewise
    \end{cases}
    \,\forall i \in observable\,buses
    \label{eq:TAM}
\end{equation}
\vspace{-.2cm}
\begin{equation}
    a_{t}=a_{t}\odot TAM
    \label{eq:maskedaction}
\end{equation}

\noindent where $a_{t}$ and $cr_{t}$ are the action and the learned criterion based on the current state $s_{t}$ and the current latent variable $c_{t}$.
The TAM is generated by comparing the voltage magnitude at each bus $i$ with the voltage criterion $cr_{t}$, as shown by (\ref{eq:TAM}).
The action is filtered by conducting a element-wise multiplication with TAM, as shown by (\ref{eq:maskedaction}).

As can be seen from the above process, at each time step, a specified voltage criterion is generated  based on the current state and the operation scenario information provided by the latent variable. 
Compared with the fixed action mask, the TAM is flexibly adjusted as the states vary, and the control actions are filtered accordingly.
In the TAM method, the physical knowledge is introduced by defining an upper bound and a lower bound for the learnable criterion, which reasonably reduces the search space and facilitates the training process.

\subsection{Physics-informed Guided Meta ES with TAM}
The complete physics-informed guided meta ES method with TAM for grid voltage control to mitigate FIDVR is shown in \textbf{Algorithm 3}.
The algorithm is composed of three major parts: 1) generate the latent variable (lines 6-8); 2) perturb the policy parameters and get the associated rewards (lines 9-20); 3) update the policy parameter and the surrogate gradient subspace (lines 21-24). In the second part, the FIDVR simulations under different perturbation directions are implemented in parallel. More specifically, as there are in total $N$ perturbation directions, and under each perturbation direction, $m$ power flow cases are run to get the associated rewards. These power flow cases are independent and can be run in parallel; the $N$ perturbation directions are also simulated in parallel, as each perturbation direction is also independent from each other. This fully parallelized framework will substantially accelerate the convergence of the algorithm, especially in the case of extremely large-scale power system simulation \cite{huang2021accelerated_DRL}. 

The major difference of the above physics-informed guided meta ES method with TAM from the guided meta ES method lies in that in the former method, the policy network will output not only the action but also the voltage criterion, which is later used to regulate the NN output (i.e., masking unnecessary actions), as shown by lines 14-16. This augmentation of the algorithm leads to a remarkable learning performance improvement with negligible extra computational efforts, since the voltage criterion $cr_{ijt-}$ only adds a few additional output dimensions to the policy network. 
In the next section, we will further test and demonstrate the superiority of the proposed physics-informed ES method through comparative studies .

\begin{algorithm}
\label{algo:guided meta ES}
 \caption{Physics-informed Guided Meta ES for Power System Control}
 \begin{algorithmic}[1]
\STATE Initialize the learning rate $\eta$, the decay rate $\xi$, the weight factor $\alpha$, the noise standard deviation $\sigma$, the total number of perturbation directions $N$, the number of surrogate gradients to use $k$, the number of top-performing directions $b$, the number of power flow cases to simulate for each iteration $m$, and the policy parameter $\theta$
\STATE Initialize the latent variable \textbf{\emph{c}} for each training power flow case
\STATE Initialize the surrogate gradient buffer $B\in\mathbb{R}^{k\times n}$
  \FOR {iteration $t$ = 1 \TO $M$}
     \STATE Sample $m$ power flow cases $\left\{\mu_{j}|j=1,...,m\right\}$
     \IF{mod($t$,\,$t_{inner}$)==0}
     \STATE Update the latent variable $c_{\mu,t}$ by maximizing $J_{\mu}$
     \ENDIF
     \STATE Generate the perturbation direction $\epsilon_{1},\epsilon_{2},...,\epsilon_{N}$:
     \STATE $\epsilon_i = \alpha\epsilon^{'} + (1-\alpha)\epsilon^{''}, \epsilon^{'}\!\sim\cal{N}$(0,$I_{n}$),\,$\epsilon^{''}\!\sim\cal{N}$(0,$I_{k}$)
     \FOR {$i$ = 1 \TO $N$}
        \STATE Generate a pair of policy parameters $\theta_{t} + \sigma \epsilon_{i}$ and $\theta_{t} - \sigma \epsilon_{i}$
        \FOR {$j$ = 1 \TO $m$}
        \STATE Generate control action and the learnt criterion $a_{ijt+}, cr_{ijt+}=\pi(s_{ijt+},c_{j,t}|\theta_{t}+\sigma\epsilon_{i})$  
        and $a_{ijt-},cr_{ijt-}=\pi(s_{ijt-},c_{j,t}|\theta_{t}-\sigma\epsilon_{i})$ 
        \STATE Generate the binary action mask vector based on the observed voltage level in $s_{ijt+}$ and $s_{ijt-}$ 
        \STATE Apply the action mask to the control actions and collect the rewards $r_{tij+}$ and $r_{tij-}$
        \ENDFOR
        \STATE Calculate the average reward:
        \STATE $r_{ti+} = \frac{1}{m}\sum_{j=1}^{m}r_{tij+},\, r_{ti-} = \frac{1}{m}\sum_{j=1}^{m}r_{tij-}$
    \ENDFOR
    \STATE Select top $b$ rewards with the largest values and update $\theta_{t}$ with the surrogate gradient:
    \STATE $g=\frac{1}{b\sigma}\sum^{b}_{i=1}(r_{ti+}-r_{ti-})\epsilon_{i}$
    \STATE $\theta_{t+1}=\theta_{t} +\eta g$
    \STATE Store $g$ to the buffer $B$ and update the surrogate gradient subspace $UU^{T}=I_{k}$, where $U\in\mathbb{R}^{n\times k}$
    \STATE Update the learning rate and the noise standard deviation with the decay rate: $\eta = \xi\eta$, $\sigma = \xi\sigma$
  \ENDFOR
\end{algorithmic} 
\end{algorithm}

\section{Case Studies}
In this section, we will first introduce the test environment for implementing and testing the proposed methods, then we will present the simulation results and comparisons with other state-of-the-art benchmark methods to demonstrate the performance of the proposed methods in terms of training efficiency, RL agent generalization capability, control performance, and optimality.

\subsection{Test environment and deployment details}
The proposed physics-informed guided meta ES-based learning framework is deployed on a local high performance computing cluster with a Linux operation system of 520 nodes. Each node has a dual-socket Intel Haswell E5-2670V3 CPU with 64 GB DDR4 memory and 12 cores per socket running at 2.3 GHz.
The training and testing of the algorithm are performed with IEEE 300-bus system \cite{huang2019_cmpldw}. 
The power system dynamic simulation is completed by the open-source platform RLGC \cite{Huang2020_DRL,RLGC}.
A summary of the hyper-parameters of the algorithm is shown in Table~\ref{table:hyper}.
Note that the policy network is constructed as a neural network with two hidden layers, one LSTM layer and one fully-connected layer, with each having 32 neurons. 
The state is defined as a vector with 154 elements, where the first 108 elements are the bus voltage magnitudes, and the last 46 elements are the remaining load levels at the buses with controllable loads.
The state vector is further concatenated with a latent context vector with 16 latent variables as the input to the policy network.
The output from the policy network is an action vector with 51 elements, where the first 46 elements are the amount of shed load, and the last 5 elements define the learnt voltage criterion, namely $V_{th,1},V_{th,2},V_{th,3},t_{1}, t_{2}$ in (\ref{eq:reward2}).
Based on the physical knowledge, an upper bound and a lower bound are defined for the above 5 fixation points as follows: $V_{th,1}\!\in$[0.7., 0.85] p.u., $V_{th,2}\!\in$[0.85., 0.92] p.u., $V_{th,3}\!\in$[0.92 0.96] p.u., $t_{1}\!\in$[0.25, 0.4] s, $t_{2}\!\in$[0.4, 0.6] s. 

In the simulation, the FIDVR events are triggered by three-phase faults at different buses which first lead to the stalling of A/C motors and eventually cause FIDVR events. To replicate the FIDVR problem, we first model the loads as single-phase induction motors (33$\%$) plus static loads (67$\%$). For dynamic modeling and parameters of the single-phase induction motors, we use the performance model and parameters recommended by NERC\cite{NERC2016}. Specially, the motor stalling time is 0.05 s and stalling voltage threshold is 0.5 p.u., which means the motors will operate in a stalled mode if the motor terminal voltage is depressed to 0.5 p.u. for a duration of more than 0.05 s.  For training the algorithm, 36 operation scenarios are generated, which combines 4 power flow scenarios with 9 fault scenarios.
The power flow scenarios vary in their generation levels and loading levels, and the fault takes places at 9 different buses. 
The fault is assumed to start at 1.0 s and ends after 0.1 s.
For testing the algorithm, 136 operation scenarios are generated, which combines 4 power flow scenarios with 34 fault scenarios.
Compared with the training cases, 25 more fault buses are considered during testing. In addition, the fault is assumed to start at 0.5s and ends after 0.08s, which is also different from the training cases.
The reason for applying new test scenarios is to validate the adaptability of the proposed data-driven control policy.
The detailed power flow conditions for training and testing are shown in Table~\ref{tab_pfcases}. The fault locations for training and testing are shown in Table~\ref{tab_faultlocation}. Each training or test scenario is simulated for 10 seconds. During the 10-second dynamic simulation, the policy network obtains the observations (voltages and percentage of remaining loads) and reward from the grid environment and provides control actions back to the grid environment every 0.1 s.

\begin{table}[t]
\caption{Power flow scenarios for training and testing}
\begin{center}
\begin{tabular}{|c|c|c|}
\hline
\textbf{Power flow scenarios} & \textbf{Generation} & \textbf{Load} \\
\hline
{\textbf{Scenario 1}} & \makecell[c]{100\% for all generation\\(22929.5 MW)}&\makecell[c]{100\% for all loads\\(22570.2 MW)}\\ 
\hline
{\textbf{Scenario 2}}& 120\% for all generators & 120\% for all loads \\ 
\hline
{\textbf{Scenario 3}}& 135\% for all generators & 135\% for all loads \\ 
\hline
{\textbf{Scenario 4}}& 115\% for all generators & 150\% for loads in Zone 1 \\ 
\hline
\end{tabular}
\label{tab_pfcases}
\end{center}
\end{table}

\begin{table}[t]
\caption{Bus indices of fault locations for training and testing}
\begin{center}
\begin{tabular}{|c|c|}
\hline
\textbf{Training} & \textbf{Testing}\\
\hline
{3,5,12,2,8,15,17,23,26}&{3,5,12,2,8,15,17,23,}\\
{}&{26,1,4,6,7,9,10,11,13,}\\
{}&{14,16,19,20,21,22,25,87,102,}\\
{}&{89,125,160,320,150,123,131,130}\\
\hline
\end{tabular}
\label{tab_faultlocation}
\end{center}
\end{table}

\begin{table}[t]
\caption{Hyperparameters for Guided Meta ES with TAM}
\begin{center}
\begin{tabular}{|l|c|c|}
\hline
\textbf{Parameters}  & \textbf{300-Bus} \\
\hline
Policy Model & LSTM+FC \\
\hline
Policy Network Size (Hidden Layers) & [32,32] \\
\hline
Weight factor ($\alpha$)& 0.5\\
\hline
Number of Disturbances ($N$)  & 128 \\
\hline
Number of surrogate gradients to use ($k$)  & 16 \\
\hline
Top Directions ($b$)  & 64 \\
\hline
Step Size ($\eta$)  & 1\\
\hline
Std. Dev. of Exploration Noise ($\sigma$)  & 2 \\
\hline
Decay Rate ($\xi$) & 0.998 \\
\hline

\end{tabular}
\label{table:hyper}
\end{center}
\vspace{-4mm}
\end{table}

\subsection{Comparison studies and performance metrics}
In this subsection, we will analyze the control performance of the proposed method for mitigating FIDVR problems and compare it with several benchmark algorithms. We define two performance metrics, namely efficiency and adaptability. For efficiency, we refer to the number of iterations for the training to converge to a reward threshold as measurements. One algorithm has a higher efficiency compared with another algorithm if it uses less training iterations to converge to the reward threshold during the training. In the following case studies, $-5\times10^3$ is used as the reward threshold. For adaptability, we refer to the average reward and the number of failed cases gained by the algorithm in unseen test scenarios as measurements. One algorithm has a better adaptability compared with another algorithm if it has higher average reward and less number of failed cases during the testing. The failed case is where the control policy fails  to  recover  the  system  voltage level  to  the  desired  criterion,  which  is  indicated  by  a  reward smaller than $-10^4$ (the large penalty set in (3) and used in this paper) .
\subsubsection{Improvement on exploration efficiency and adaptability through guided search}
\begin{figure}[t]
\centerline{\includegraphics[scale=0.3]{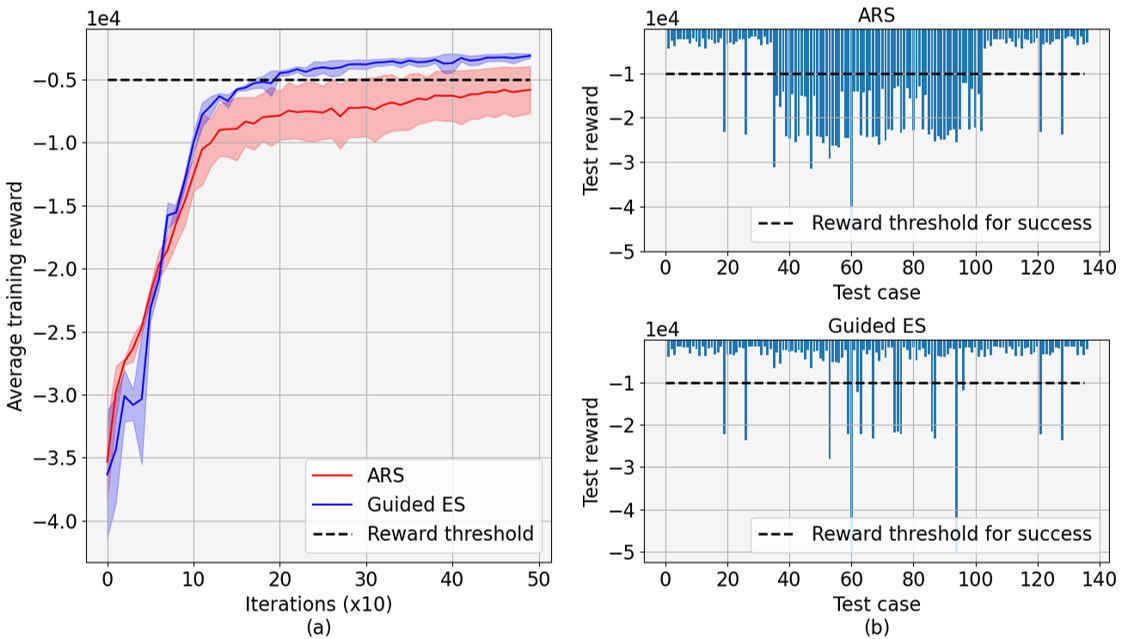}}
\caption{Comparison of ARS with guided ES:(a) average training reward over 3 random seeds; (b) test reward for 136 new cases.}
\label{fig:compareARSwithguidedES}
\end{figure}

We first compare the performance of the guided ES method with the ES method to evaluate the function of the surrogate gradient in terms of exploration efficiency and adaptability. 
Fig.~\ref{fig:compareARSwithguidedES} presents the training results and the testing results for both methods. 
For the ES method, we applied the ARS from our previous work \cite{huang2021accelerated_DRL}, which is an improved version of ES.

Fig.~\ref{fig:compareARSwithguidedES}(a) shows the average reward for 500 training iterations, where the shaded area stands for the standard deviation over 3 random seeds.
As can be seen from the figure, during the training, the reward curve of the guided ES method converges faster than that of the ARS method by demonstrating a larger slope from 0 to 100 iterations, and it also reaches a higher converged value with a smaller deviation range at the end of the training.
As shown in the figure, on average, the ARS method did not achieve the reward threshold after 500 iterations, while the guided ES method reaches the threshold only after 200 iterations. This is because the ARS method applies a complete random search in the action space, which brings in additional exploration efforts. On the contrary, the guided ES leverages the guidance from the surrogate gradient such that it can explore more in those promising directions and thus achieve more effective exploration, as illustrated in Fig. \ref{fig:ESillustration}, which greatly improves exploration and sample efficiency. 
Fig.~\ref{fig:compareARSwithguidedES}(b) shows the reward gained in each of the 136 test cases from the two methods. 
Table~\ref{table:testresults} lists the average test reward for the two methods. 
As shown in the table, the guided ES method improved the average reward by more than 50\% compared with the ARS method.
We further compared the number of failed cases of the two methods.
As shown in the third column of Table~\ref{table:testresults}, guided search helps reduce the number of failed cases by 75\% compared with the ARS method.
Therefore, we can safely conclude that the guided search based on the surrogate gradient also contributes to more adaptive control policies.

\begin{table}[t]
\caption{Comparison of test results}
\begin{center}
\begin{tabular}{|c|c|c|}
\hline
\textbf{Method} & \textbf{Average test reward}&\textbf{No. of failed cases} \\
\hline
ARS & $-1.27\times10^4$&72 \\
\hline
Guided ES &$-5.6\times10^3$&17 \\
\hline
Guided meta ES &$-4.3\times10^3$&12\\
\hline
Guided meta ES + mask &$-2.8\times10^3$&8 \\
\hline
Guided meta ES + TAM &$-1.89\times10^3$&3 \\
\hline
MPC &$-1.82\times10^3$&3 \\
\hline
\end{tabular}
\label{table:testresults}
\end{center}
\vspace{-4mm}
\end{table}

\subsubsection{Meta learning for adaptability enhancement}
\begin{figure}[t]
\centerline{\includegraphics[scale=0.23]{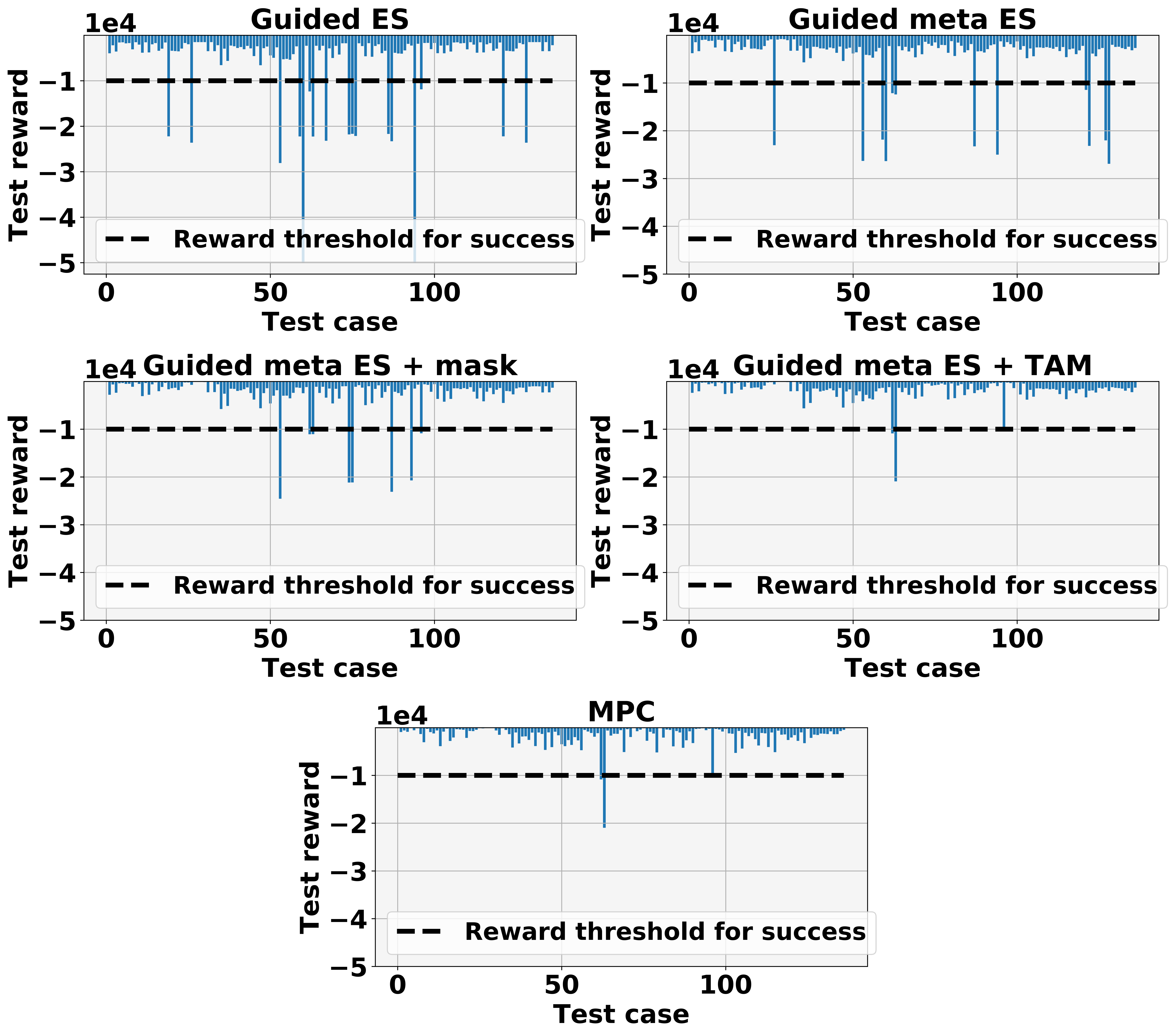}}
\caption{Test results of the proposed methods and the benchmark methods}
\label{fig:compareguidedEStest}
\end{figure}

The proposed physics-informed guided meta ES with TAM method is compared with three other benchmark guided search methods, namely the guided ES method, the guided meta ES method, and the guided meta ES method with mask derived from the fixed voltage criterion.  Fig.~\ref{fig:compareguidedES} shows the training curves of the four methods, and Fig.~\ref{fig:compareguidedEStest} shows the test results of the four methods. The test results are also listed in Table~\ref{table:testresults}.

We first look into the function of meta learning for enhancing the adaptability of the control policy. In Fig.~\ref{fig:compareguidedES}, the training curve of the guided ES method and the guided meta ES method show a similar trend. However, by comparing the test rewards in Fig.~\ref{fig:compareguidedEStest} and Table~\ref{table:testresults}, it can be discovered that the guided meta ES method can achieve a higher average test reward with fewer failed cases. This is because the implemented MSO strategy within the guided meta ES method can quickly fine-tune the learnt control policy based on the newly extracted features from the unseen test cases, leading to a more adaptive control performance.

\subsubsection{Boosting efficiency and adaptability with physics-informed action mask}
\begin{figure}[t]
\centerline{\includegraphics[scale=0.35]{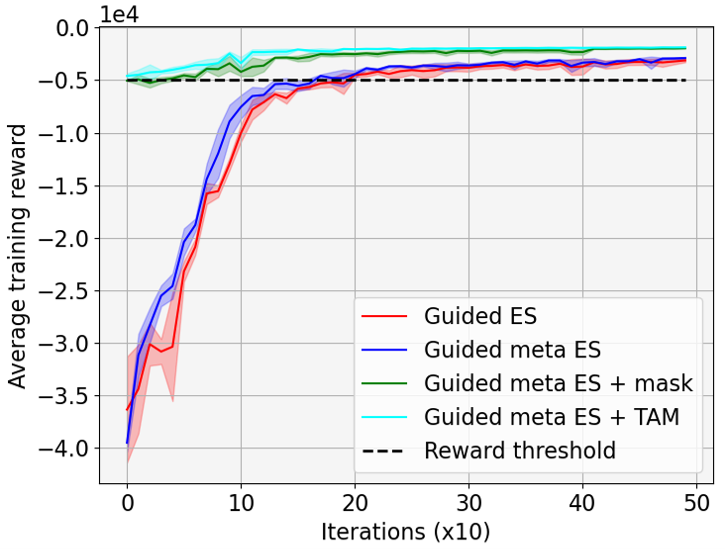}}
\caption{Comparison of training curves of guided ES methods}
\label{fig:compareguidedES}
\end{figure}

We further analyze the function of action mask by comparing the training and test rewards among the four methods.
As shown in Fig.~\ref{fig:compareguidedES}, the last two methods with an action mask outperform the first two methods without an action mask with a much higher starting point and also a higher final reward. The last two methods reach the reward threshold at the very beginning of the training, while it takes around 150 iterations for the first two methods to reach the same threshold. There is at least 80\% improvement in terms of sample efficiency. Therefore, it can be safely concluded that the embedding of physics knowledge through the action mask can considerably boost the exploration efficiency of the algorithm. 

With respect to adaptability, based on Table~\ref{table:testresults}, it can be observed that the last two ES methods with an action mask have higher average test rewards and fewer failed cases than the other ES methods without an action mask. This verifies the contribution of the action mask in enhancing the adaptability of the learnt control policy. In addition, it should be noted that the proposed guided meta ES with TAM outperforms all the other methods with the highest average test reward and the fewest failed case. Thus, a learnt voltage criterion from TAM that flexibly adjusts based on the environment works more efficiently in generating masks for action selection than a fixed voltage criterion.


\begin{figure}[t]
\centerline{\includegraphics[scale=0.23]{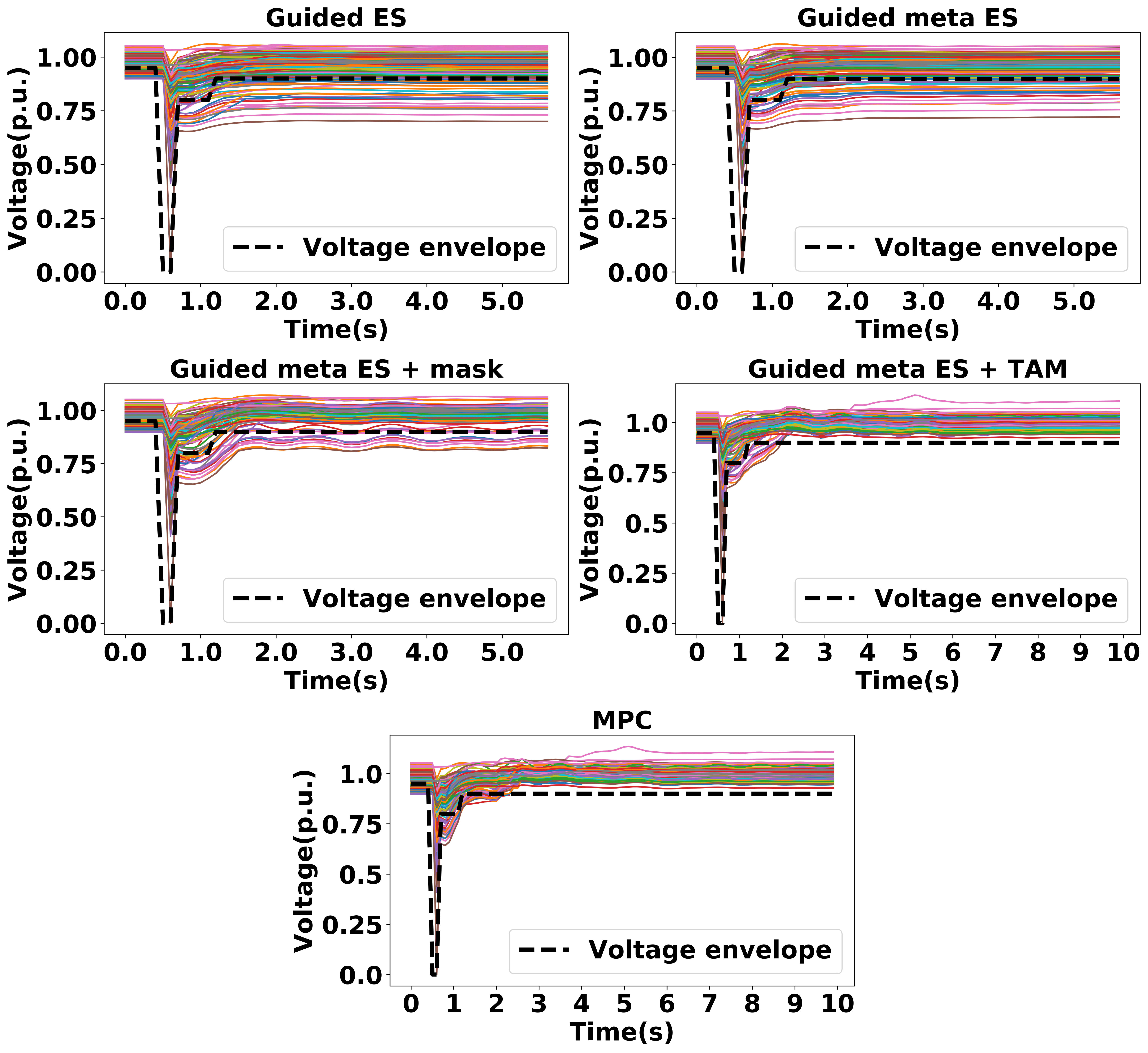}}
\caption{Comparison of bus voltage under the four control methods}
\label{fig:comparevoltage}
\end{figure}

To look deeper into how the TAM helps improve the learning performance, we show one test case in which the first three methods failed and only the guided meta ES with TAM succeeded in restoring the voltage level.
The voltage magnitudes of all observable buses under the four methods are shown in Fig.~\ref{fig:comparevoltage}.
In total there are 108 voltage curves in each figure, representing the voltage magnitudes at 108 observable buses in the system.
The dashed black voltage envelope stands for the lower security bound of the voltage. 
For the first three methods, the simulation ends at around 6 seconds. 
This is because the control policies failed to restore the system voltage to 0.95 p.u. after 5 seconds when the fault is cleared. The simulation thus terminates in advance.
For the proposed guided meta ES with TAM method, the simulation lasts for 10 seconds, which is the predefined time length of an entire simulation, and all the bus voltage magnitudes are above the voltage recovery criterion envelope.

\begin{figure}[t]
\centerline{\includegraphics[scale=0.23]{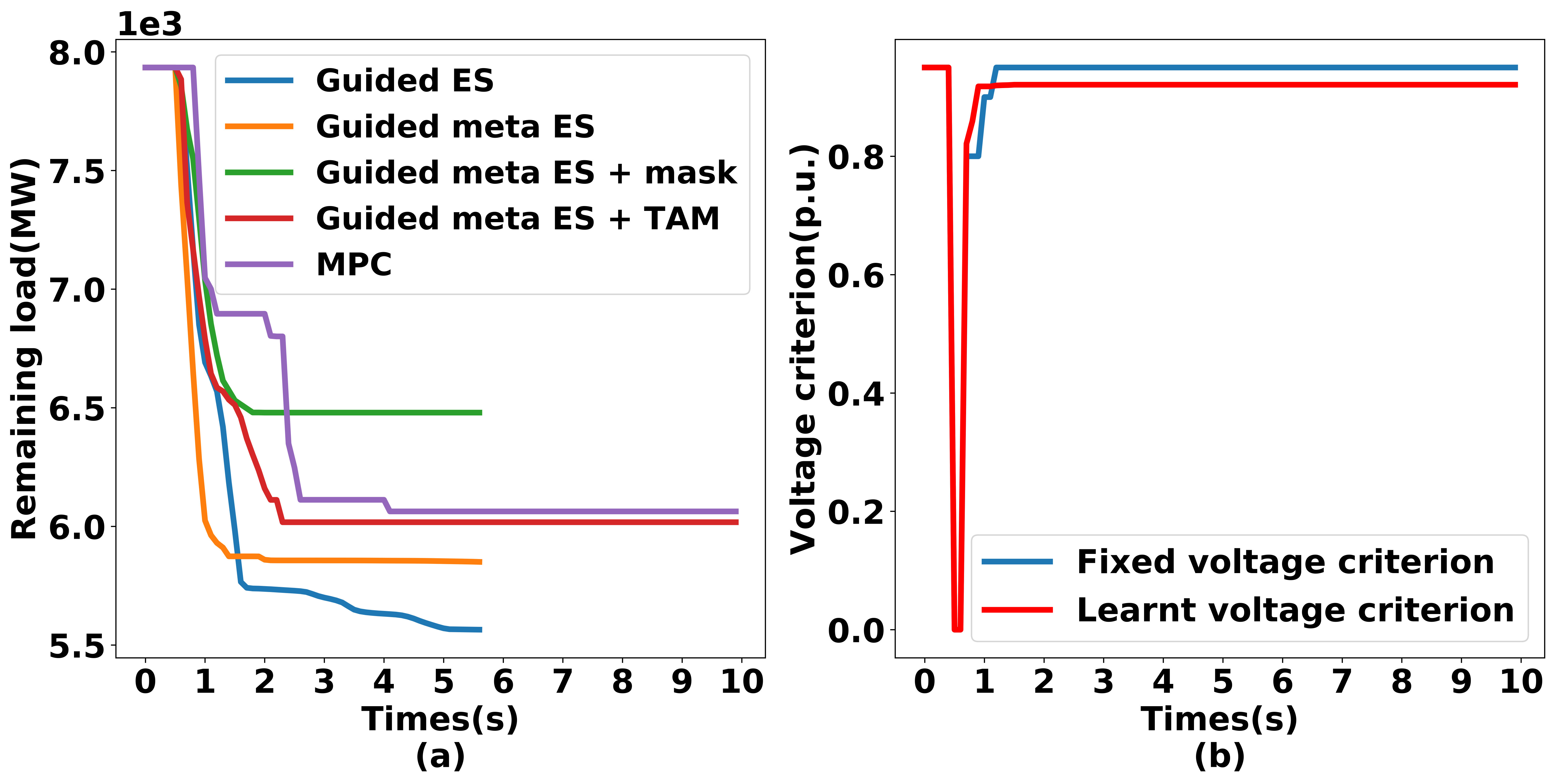}}
\caption{Comparison of load shedding strategy and mask voltage criterion: (a) total remaining load of the system; (b) voltage criterion.}
\label{fig:compareaction}
\centering
\end{figure}

To understand why the voltage recovery strategy from our proposed method is more effective than the other four benchmark guided search methods, we further look under the hood by comparing the load shedding actions generated by the four methods, as shown in Fig.~\ref{fig:compareaction}(a). The figure shows the total remaining load for all tested methods. A higher remaining load level indicates less load interruptions and is thus more desired. As shown in the figure, the first two methods have relatively lower remaining load, which implies that unnecessary load shedding is conducted without guidance from the physical knowledge, since they did not apply the mask technique for filtering actions.
The third method, which utilizes a mask from a fixed voltage criterion for action filtering, has the highest remaining load level after shedding.
However, the system voltage cannot be fully recovered in this case due to insufficient load shedding, which deviates from the initial goal of FIDVR mitigation and voltage recovery. Our proposed guided meta ES with TAM method achieves a remaining load level in between the above methods, which indicates that it has achieved a balanced trade-off: meeting the voltage recovery goal with minimal load interruptions.

Fig.~\ref{fig:compareaction}(b) compares the learnt voltage criterion from TAM and the fixed voltage criterion. 
In TAM, for each time step, a voltage criterion is generated. 
For a complete simulation with a time length of 10s and a control time step of 0.1s, there are 100 learnt voltage criteria.
We compare the average value of the learnt voltage criteria with the fixed voltage criterion.
As can be seen from the figure, following the immediate occurrence of the fault (between 0.6s and 1s), the learnt voltage criterion is higher than the fixed voltage criterion, which explains why a larger amount of load is shed in the guided meta ES method with TAM. The algorithm increases the voltage bar immediately after the fault occurs to avoid the potential future voltage failures. In Fig.~\ref{fig:comparevoltage}, comparing the two figures at the second row, it can be observed that with TAM, the voltage rises faster and higher than that with a fixed mask after the fault takes place (from 0.6 s to 1 s), due to the larger amount of load shedding.
Therefore, we can safely conclude that a learnt voltage criterion from TAM can lead to more reliable and adaptive load shedding strategies.
\subsubsection{Computation time for implementation}
We also compare the computational time of the four methods for determining control actions, and the results are shown in Table~\ref{table:testtime}.
The FIDVR event duration for each testing scenario is 10 s. As discussed in Section V.A, all the four methods obtain the observations from the grid simulation environment and provide control actions back to it every 0.1 s. Thus, there are a total of 100 control steps with 0.1 s time step for each testing scenarios. The “average solution time” listed in the Table {\ref{table:testtime}} is the average “total computation time” for the whole 100 control steps for each testing scenario. That is, for example the proposed guided meta ES with TAM method  take 0.0063 s on average to compute control actions within the 0.1 s control interval. Therefore, our proposed method meets the real-time operation requirement and is suitable for fast, short-term voltage recovery.

\begin{table}[t]
\caption{Comparison of computational time}
\begin{center}
\begin{tabular}{|c|c|}
\hline
\textbf{Method}  & \textbf{Average solution time (s)}\\
\hline
Guided ES & 0.51\\
\hline
Guided meta ES &0.49\\
\hline
Guided meta ES + mask & 0.62\\
\hline
Guided meta ES + TAM & 0.63\\
\hline
MPC & 64.34\\
\hline
\end{tabular}
\label{table:testtime}
\end{center}
\vspace{-4mm}
\end{table}

\subsubsection{Comparison with MPC-based voltage control method}
Finally, to further demonstrate the effectiveness of our method, we compare it with the MPC  method that has been used for FIDVR mitigation control in the literature. The test results of the MPC method are shown in the last row of Table~\ref{table:testresults} and also Fig.~\ref{fig:compareguidedEStest}. As shown in the table and the figure, the guided meta ES method with TAM and the MPC method have the same number of failed cases, and the former achieved an average test reward very close to the latter.
The voltage magnitudes under the two methods can be further compared in Fig.~\ref{fig:comparevoltage}. As can be seen from the figure, the voltage profiles obtained with both methods are very similar. The same is true for the remaining load levels (or total load shedding amounts). The MPC method relies on complete knowledge and models of the power grid environment, hence theoretically it can reach a (near) global optimal solution. However, the assumption of accurate modeling of the power grid environment is untenable in reality, especially for large-scale power systems with extreme complexities. Furthermore, the computational burden of the MPC method makes it impractical for real-time fast voltage control. As shown in Table~\ref{table:testtime}, the solution time of the MPC method is 100 times higher than that of the proposed guided meta ES method with TAM, which does not meet the “real-time” requirements for the load shedding actions. This comparison study shows that our method can quickly respond to the  emergencies that are unseen during training with a near-optimal voltage recovery policy, indicating its good potential for real system implementations.

\section{Conclusions}
In this paper, we propose a physics-informed guided meta ES algorithm to mitigate the FIDVR problem.
The proposed algorithm makes use of guidance from physics knowledge via the TAM technique and the surrogate gradient to conduct more efficient exploration. The physics-embedding also help achieve more robust solutions, particularly for unseen scenarios.
In addition, the algorithm is combined with meta-learning to gain adaptability.
Simulation results on the IEEE 300-bus system show that the proposed algorithm outperforms other state-of-the-art model-free control algorithms with a faster convergence during training (i.e., 80\% sample efficiency improvement over the basic ES method) and a more adaptive control strategy to unseen fault scenarios (i.e., reducing the number of failed cases from 72 to 3 out of 136 test cases) , and also meets the real-time requirement with 0.006 s solution time per control step.

For future research, given the good results for the FIDVR problem, applying the developed method to other power system control applications are promising and worthwhile. Also, trade-off between the guided direction and the random search in the guided ES method for better learning performance deserves more investigations. Last but not least, it is of great interest to study the transferability of the learnt control policy among different test systems, which could lead to more general and practical algorithm implementation.


%






\bibliographystyle{IEEEtran}
\bibliography{references}

%








\end{document}